%% file: main.tex
\documentclass[conference]{IEEEtran}
\IEEEoverridecommandlockouts

\usepackage{stmaryrd}
\usepackage{amsmath}
\usepackage{amssymb}
\usepackage{mathtools}
\usepackage{graphicx}
\usepackage{tikz}
\usetikzlibrary{quantikz2}
\usetikzlibrary{automata, positioning}
\usepackage{braket}
\usepackage{mathrsfs} 
\usepackage{flushend}
\usepackage{mathdots}

\usepackage{amsmath}
\usepackage{amsfonts}%
\usepackage{amssymb}
\usepackage{amsthm}
\usepackage{mathtools}
\usepackage{algorithmicx}
\usepackage{algpseudocode}
\usepackage{algorithm}
\usepackage{physics}
\usepackage{relsize}
\usepackage{gensymb}
\usepackage{stmaryrd}
\algdef{SE}[DOWHILE]{Do}{doWhile}{\algorithmicdo}[1]{\algorithmicwhile\ #1}%

\usepackage{cite}  
\usepackage{array}
\usepackage{subcaption}
\usepackage[font=normalsize,labelformat=simple]{subfig}

\usepackage{blindtext}
\usepackage{tikz}
\usepackage{amsthm}
\newtheorem{lemma}{Lemma}

\newtheorem{example}{Example}

\newcommand{\supp}{\mathrm{supp}}
\newcommand{\setv}{\mathcal{V}}
\newcommand{\setc}{\mathcal{C}}
\newcommand{\sete}{\mathcal{E}}

\title{Turbo-Annihilation of Hook Errors in Stabilizer Measurement Circuits}
\author{\IEEEauthorblockN{Michele Pacenti,
Asit K. Pradhan, Shantom K. Borah
and
Bane Vasi\'c}
\IEEEauthorblockA{Department of Electrical and Computer Engineering, University of Arizona, Tucson, AZ, USA}

\{mpacenti, asitpradhan, shantomborah\}@arizona.edu, vasic@ece.arizona.edu 

\thanks{The authors wish to thank Nithin Raveendran and Dimitris Chytas for the insightful discussions around this paper. We acknowledge the support of the National Science Foundation under grants ERC-1941583, CIF-2420424, CIF-2106189, CCF-2100013, CCSS-2052751, and a generous gift from our friends and Maecenases Dora and Barry Bursey. Bane Vasi\'{c} has disclosed an outside
interest in his startup company Codelucida to The University
of Arizona. Conflicts of interest resulting from this interest are being managed by The University of Arizona in accordance
with its policies.

}

}

\begin{document}

\maketitle

\begin{abstract}
We propose a scalable decoding framework for correcting correlated hook errors in stabilizer measurement circuits. Traditional circuit-level decoding attempts to estimate the precise location of faults by constructing an extended Tanner graph that includes every possible source of noise. However, this results in a highly irregular graph with many short cycles, leading to poor performance of message-passing algorithms. To compensate, ordered statistics decoding is typically employed, but its cubic complexity renders it impractical for large codes or repeated stabilizer measurements. Our approach instead focuses on estimating the effective data errors caused by hook faults, modeling them as memory channels. We integrate trellis-based soft-input soft-output equalizers into the Tanner graph of the code, and show that the resulting decoding graph preserves the structural properties of the original Tanner graph such as node degree and girth, enabling efficient message passing. Applied to bivariate bicycle quantum LDPC codes, our decoder outperforms standard belief propagation on the circuit-level graph and closely approaches OSD0 performance, all while maintaining linear complexity and scalability.
\end{abstract}

\section{Introduction}
The effect of various noise sources in quantum memories can be modeled and corrected using the circuit-level Tanner graph framework \cite{higgott_improved_2023}. In this model, each noise source in the circuit corresponds to a variable node, while check nodes represent the outcomes of ancilla measurements. An edge connects a variable node to a check node if the corresponding noise source can trigger that syndrome measurement. Noise sources that lead to the same syndrome are merged into a single variable node. Given a measured syndrome, the decoder operates on this graph to infer the most likely set of faults.
However, this approach presents several limitations. First, the circuit-level Tanner graph is significantly larger than the Tanner graph of the underlying code, particularly when multiple rounds of measurement are involved. This increase arises from the decoder’s attempt to identify the exact location of each fault (up to column merging), rather than estimating their collective effect on the data qubits. Second, the circuit-level graph does not retain the desirable structural properties of the original code graph, thus vanishing any effort in the QLDPC design optimization. In particular, it typically exhibits irregular variable and check degrees, including many variable nodes of degree one, and contains a large number of short cycles \cite{gong2024lowlatencyiterativedecodingqldpc}. As a result, standard message-passing algorithms often perform poorly when applied to the circuit-level Tanner graph. To recover acceptable performance, it is usually necessary to employ Ordered Statistics Decoding (OSD), which increases the decoding complexity to $\mathcal{O}(N^3)$, where $N$ is the number of variable nodes in the circuit-level graph. Since this number is typically much larger than in the original Tanner graph, the approach becomes impractical for large codes or for repeated syndrome extraction rounds.

More recently, several approaches have been proposed to decode circuit-level noise, as alternatives to BPOSD. We highlight some of the most noticeable. Kuo and Lai \cite{kuo2024faulttolerantbeliefpropagationpractical} propose a belief-propagation decoder operating on a mixed-alphabet circuit-level parity check matrix, achieving an asymptotic complexity of $\mathcal{O}(N)$ for a single measurement round, where $N$ is the blocklength of the circuit-level matrix, which is is $N = n+mw+2m$, where $m$ is the number of checks and $w$ the stabilizer weight. They propose two versions, one operating over $\mathrm{GF}(4)$, and the other over $\mathrm{GF}(16)$, both utilizing serial scheduling and adaptive parameter tuning. Despite the linear complexity, the authors adopt the technique of \textit{diversity decoding}, which consists of running several decoders in parallel (in the specific case around 60), with slightly different parameters or update rules, to correct the same error; moreover, they utilize a serial scheduling, meaning that the variable and check nodes are updated one-by-one, rather than all in parallel. Although this is done to optimize the performance, diversity decoding demands a high amount of parallel computational resources, while serial scheduling notoriously introduce more decoding latency than the parallel scheduling.  Gong \textit{et al.} \cite{gong2024lowlatencyiterativedecodingqldpc} propose a windowed guided decimation BP decoder operating on the circuit-level Tanner graph, which has a complexity of $\mathcal{O}(N^2)$, achieving performance comparable to the one of the standard BPOSD, but with reduced decoding latency. Wolanski and Barber \cite{wolanski2025ambiguityclusteringaccurateefficient} proposed the ambiguity clustering decoder which performs a partial OSD processing, thus reducing the decoding complexity. These contributions aim to reduce the complexity of the BPOSD, but they all operate on variants of the circuit-level Tanner graph.

In our approach, we aim to decode the effective data error resulting from the collective effect of all faults, rather than determining their precise locations. At the same time, we account for the correlated nature of the resulting noise. To do this, we draw inspiration from the turbo equalization technique \cite{turbo_eq}, widely used in classical communication systems affected by channels with memory, such as the inter-symbol interference (ISI) channel. In such channels, limited bandwidth causes transmitted pulses to spread and overlap, introducing correlations between neighboring bits. This form of correlated noise can be modeled as a finite-state machine (FSM), and equalized using a Viterbi detector \cite{viterbi}. In turbo equalization, a soft-input soft-output estimator such as the Bahl-Cocke-Jelinek-Raviv (BCJR) algorithm \cite{bcjr} replaces the Viterbi detector and iteratively exchanges soft information with a belief propagation (BP) decoder. Other approaches model the effect of ISI as a second Tanner graph, and turbo equalization corresponds to joint decoding over both the code and channel graphs \cite{siegel_isi, siegel_isi2}.

We propose a new decoding framework for QLDPC codes that accounts for correlated errors introduced by circuit-level noise. In particular, we focus on a type of correlated fault known as a hook error. A hook error originates from an ancilla qubit and propagates to one or more data qubits involved in the same stabilizer measurement. To isolate the effect of hook errors, we consider a simplified scenario in which noisy $X$ stabilizer measurements are followed by perfect $Z$ stabilizer measurements, all within a single round of syndrome extraction. The perfect $Z$ measurements allow correction of $X$ errors caused both by single-qubit depolarizing noise and by faulty CNOT gates.
To model the impact of hook errors, we introduce an auxiliary Tanner graph in which circuit faults are represented as variable nodes and data qubits as check nodes. This graph inherently contains numerous unavoidable four-cycles. To mitigate this, we group variable nodes corresponding to faults on the same ancilla into higher-order structures we call \textit{equalizer nodes}. These nodes capture the finite-state machine (FSM) dynamics of hook errors and are decoded locally using the BCJR algorithm. The resulting joint Tanner graph retains the essential structural properties of the original code graph while circumventing the irregular node degrees and short cycles that typically hinder circuit-level Tanner graphs.
We then develop a message-passing decoder operating on this joint graph, with computational complexity $\mathcal{O}(n)$, where $n<N$ is the blocklength of the original code. Our numerical results show that this decoder outperforms standard belief propagation (without OSD) on the circuit-level Tanner graph and approaches the performance of the BPOSD0 decoder in the low error probability regime.

The remainder of this paper is organized as follow: in Section \ref{sec:preliminary} we introduce the preliminary concepts. In Section \ref{sec:modeling} we present the construction of the joint Tanner graph. In Section \ref{sec:decoder} we illustrate the decoding algorithm, and in Section \ref{sec:results} we present numerical results.

\section{Preliminaries}
\label{sec:preliminary}
\subsection{Low-density parity check codes and Tanner graphs}
An $[n,k,d]$ linear code $C \subset \mathbb{F}_2^{n}$ is a linear subspace of $\mathbb{F}_2^{n}$ generated by $k$ elements, such that each element in $C$ has Hamming weight at least $d$. A code $C$ can be represented by an $(n-k) \times n$ parity check matrix $\mathbf{H}$ such that $C = \ker \mathbf{H}$. If $\mathbf{H}$ is \textit{sparse}, \textit{i.e.}, its row and column weights are constant with $n$, the code $C$ is a \textit{low-density parity check} (LDPC) code. The parity check matrix is the biadjacency matrix of the \textit{Tanner~graph} $\mathcal{T}=(\setv\cup \setc,\sete)$, where the nodes in $\setv$ are called \textit{variable nodes} and the nodes in $\setc$ are called \textit{check nodes}, such that and there is an edge between $v_j \in \setv$ and $c_i \in \setc$ if $h_{ij}=1$, where $h_{ij}$ is the element in the $i$-th row and $j$-th column of $\mathbf{H}$. The \textit{degree} of a node is the number of incident edges to that node. If all the variable (check) nodes have the same degree we say the code has \textit{regular} variable (check) degree, and we denote it with $\gamma$ ($\rho$). We denote a biregular code as $(\gamma,\rho)$-regular. A \textit{cycle} is a closed path in the Tanner graph, and we denote its length by the number of variable and check nodes in the cycle. The \textit{girth} $g$ of a Tanner graph is the length of its shortest cycle.

\subsection{Calderbank-Shor-Steane codes}
Let $(\mathbb{C}^2)^{\otimes n}$ be the $n$-dimensional Hilbert space, and $P_n$ be the $n$-qubit Pauli group; a \textit{stabilizer} group is an Abelian subgroup $S \subset P_n$, and an $\llbracket n,k,d \rrbracket$ stabilizer code  is a $2^k$-dimensional subspace $\mathcal{C}$ of $(\mathbb{C}^2)^{\otimes n}$ that satisfies the condition $S_i\ket{\Psi} = \ket{\Psi},\ \forall\ S_i\in S, \ket{\Psi}\in \mathcal{C}$. An $\llbracket n, k_X-k_Z^{\perp}, d \rrbracket $ Calderbank-Shor-Steane (CSS) code $\mathcal{C}$ is a stabilizer code constructed using two classical  $[n,k_X,d_X]$ and $[n,k_Z,d_Z]$ codes $C_X = \ker \mathbf{H}_X$ and $C_Z = \ker \mathbf{H}_Z$, respectively, such that $C_Z^{\perp} \subset C_X$ and $C_X^{\perp} \subset C_Z$ \cite{calderbank_good_1996}. The minimum distance is $d\geq~\mathrm{min}\{d_X,d_Z\}$, with $d_X$ being the minimum Hamming weight of a codeword in $C_X \setminus C_Z^{\perp}$, and $d_Z$ being the minimum Hamming weight of a codeword in $C_Z \setminus C_X^{\perp}$. A quantum LDPC (QLDPC) code is a CSS code where both $\mathbf H_X$ and $\mathbf{H}_Z$ are sparse.

\subsection{Syndrome extraction circuit}

To a Pauli operator $E$ is associated a syndrome $\mathbf{s} \in \mathbb F_2^{n-k}$, such that
\begin{equation}
    \begin{cases}
        s_i=0\ \mathrm{if}\ E\ \mathrm{commutes\ with}\ S_i \\
        s_i=1\ \mathrm{otherwise}.
    \end{cases}
\end{equation}
Each stabilizer $S_i \in \mathcal{S}$ is measured using an ancillary qubit and a series of controlled operations. Let $S_{i,j}$ be the Pauli operator acting on the $j$-th qubit in $S_i$. The measurement circuit consists of three main steps.
First, the ancillary qubits are initialized according to the stabilizer type. Ancillas used for the measurement of $X$-stabilizers are prepared in the $|+\rangle$ state, while those used for $Z$-stabilizers are prepared in the $|0\rangle$ state.
Next, a sequence of controlled-NOT (CNOT) operations entangles the ancilla with the data qubits involved in the stabilizer measurement. The set of participating qubits is given by  
$$
J_X = \{j \mid S_{i,j} \in \{X, Y\} \}, \quad J_Z = \{j \mid S_{i,j} \in \{Z, Y\} \}.
$$
For $X$-stabilizers, the ancilla serves as the control, and the data qubits in $J_X$ are the targets. Conversely, for $Z$-stabilizers, the data qubits in $J_Z$ serve as controls, and the ancilla is the target.
Finally, the ancilla is measured in the computational $X$-basis for $X$ stabilizers, and $Z$-basis for $Z$ stabilizers, producing the stabilizer syndrome bit $s_i$.

In the circuit-level noise model, qubit preparation, gates, and measurements are considered noisy. A faulty qubit preparation prepares a single-qubit state orthogonal to the correct one with probability $p$. A faulty CNOT is an ideal CNOT followed by one of the 15 non-identity Pauli errors on the control and target qubits, picked uniformly at random with probability $p/15$. A faulty measurement is an ideal measurement followed by a classical bit-flip error applied to the measurement outcome with probability $p$. Moreover, idling qubits suffer from depolarizing noise. 

\begin{figure}
    \centering
\input{tikz/circuit1.tikz}
    \caption{Possible locations of a hook error on a qubit $a$ during an $X$ stabilizer measurement. We do not consider errors after the last CNOT as they are not detectable.}
    \label{fig:hook_error}
\end{figure}
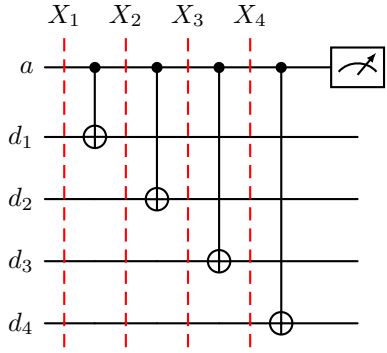

In this paper, we are mainly interested in faulty CNOT operations. Thus, we do not consider idling noise or measurement errors. In particular, we focus on the effect of \textit{hook errors}. A hook error occurs when an error on an ancillary qubit propagates to all or a subset of the data qubits involved in the measurement of the stabilizer associated with that ancilla. An example is illustrated in Fig.~\ref{fig:hook_error}, where $X$ errors may happen on the ancilla qubit $a$ at different time steps.
To isolate the impact of hook errors, we consider the following simplified experiment, illustrated in Fig.~\ref{fig:experiment}. We assume depolarizing noise on $X$ ancilla qubits and data qubits at the beginning of the circuit. We first measure all the $X$-stabilizers of a QLDPC code using noisy CNOT gates, followed by the measurement of all $Z$-stabilizers using perfect CNOT gates. We do not include idling noise or measurement errors. By leveraging the perfect $Z$-stabilizer measurements, we can correct $X$ errors on the data qubits originating from the noisy stabilizer measurements. Therefore, throughout the paper we will consider only $X$ errors, although a specular analysis holds for $Z$ errors.

\subsection{Bivariate bicycle codes}

For our experiment, we consider bicycle bivariate (BB) codes \cite{bravyi2024high}, although our analysis extends naturally to other classes of QLDPC codes. BB codes are $(\gamma,\rho)$-regular, quasi-cyclic codes with parity-check matrices given by

\begin{equation} \mathbf{H}_X = [\mathbf{A}\ \mathbf{B}],\quad \mathbf{H}_Z = [\mathbf{B}^T\ \mathbf{A}^T]. \end{equation}

Since $\mathbf{A}$ and $\mathbf{B}$ are circulant matrices, they can be compactly represented using polynomial notation as

\begin{equation} H_X(x) = [ a(x),\ b(x)],\  H_Z(x) = [ b^{*}(x),\ a^{*}(x) ], \end{equation}

where $a^*(x)$ and $b^*(x)$ denote the conjugate transposes of $a(x)$ and $b(x)$, respectively. Without loss of generality, for illustration purposes, we consider a $(2,4)$-regular bicycle code, where the circulant polynomials take the form

\begin{equation} a(x) = x^{a_1} + x^{a_2}, \quad b(x) = x^{b_1} + x^{b_2}. \end{equation}

Substituting these into the parity-check matrices, we obtain

\begin{align*} H_X(x) &= [ x^{a_1} + x^{a_2},\ x^{b_1} + x^{b_2} ]\\  H_Z(x) &= [ x^{-b_1} + x^{-b_2},\ x^{-a_1} + x^{-a_2} ]. \end{align*}

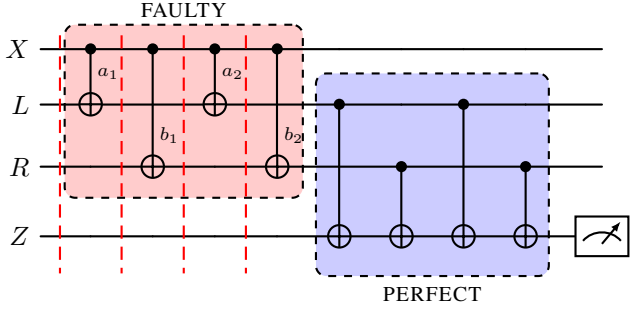
\begin{figure}
    \centering
    \input{tikz/experiment.tikz}
    \caption{Experiment setup. Each wire represents a set of $N$ qubits: $X$ is the set of $X$ ancilla qubits, $Z$ is the set of $Z$ ancilla qubits, while $L$ and $R$ are the set of left and right data qubits in BB codes. Each CNOT gate represented in the circuit is a set of parallel CNOT gates, applied according to the polynomials $a(x)$ and $b(x)$. The slices indicate error locations at different time steps.}
    \label{fig:experiment}
\end{figure}

Notice that we can distinguish two sets of qubits: the left qubits, denoted by $L$, which interact with the $X$-stabilizer checks via $\mathbf{A}$ and with the $Z$-stabilizer checks via $\mathbf{B}^T$, and
the right qubits, denoted by $R$, which interact with the $X$-stabilizer checks via $\mathbf{B}$ and with the $Z$-stabilizer checks via $\mathbf{A}^T$.
The experimental setup is illustrated in Fig.~\ref{fig:experiment}\footnote{We acknowledge $\mathtt{quantikz}$ for the drawing of the quantum circuits \cite{kay2023tutorialquantikzpackage}.}.
Since all CNOT gates corresponding to a given polynomial entry can be applied in a single time step, we represent each such set of operations as a single CNOT gate labeled with the corresponding power of $x$. As a consequence, each wire in Fig.~\ref{fig:experiment} represents a set of $N$ qubits, where $N$ is the circulant size.

\subsection{Finite state machines and trellis diagrams}

A finite-state machine (FSM) is a mathematical model for describing discrete systems with memory. At each time step $t$, the FSM receives an input $f_t$ and produces an output $x_t$, where the output depends on both the current input and the current state $\sigma_t \in \Sigma$, where $\Sigma$ is the state space. The system then transitions to a new state $\sigma_{t+1}$, determined by $\sigma_t$ and $x_t$. An FSM is can be represented by its \textit{state diagram}, where vertices represent states, and arrows represent state transitions. Each arrow is labeled as $./.$, where the left entry is the input symbol and the right entry is the corresponding output.
The temporal evolution of an FSM can be represented by a trellis diagram, which unwraps the state diagram over time. A trellis is a time-indexed directed graph, where each time slice $t$ consists of two consecutive state sets, $\Sigma_t$ and $\Sigma_{t+1}$, connected by directed edges $(\sigma, \sigma')$ representing state transitions. Each vertex in $\Sigma_t$ corresponds to a possible FSM state, and the edges retain the labeling convention of the state diagram.

\subsection{Bahl-Cocke-Jelinek-Raviv algorithm}
The BCJR algorithm \cite{bcjr} is a \textit{maximum-a-posteriori} (MAP), soft-input soft-output (SISO) algorithm operating on the trellis of a FSM.
The objective of the BCJR algorithm is to compute the MAP probabilities of a sequence of symbols $\mathbf x = (x_1,..,x_{\rho})$, namely $P(x_t|L(x_1),...,L(x_{\rho}))$, where $L(x_t)$ is the prior \textit{log-likelihood ratio} for $x_t$, defined as
\begin{equation}
     L(x_t) = \ln \left[\frac{P(x_t=0)}{P(x_t=1)} \right].
\end{equation}
The output of the algorithm is the sequence of LLRs of $\mathbf{x}$ associated to the \textit{a-posteriori} probabilities $L'(x_t)$, defined as:
\begin{equation}
    L'(x_t) = \log \frac{P(x_t=0|L(x_1),...,L(x_{\rho}))}{P(x_t=1|L(x_1),...,L(x_{\rho}))},
\end{equation}
The computation involves three quantities in logarithmic format: \textit{forward metric} $\log \alpha_t$, \textit{backward metric} $\log \beta_t$, and the \textit{branch metric} $\log \gamma_t$.
It can be shown that the logarithmic branch metric is equal to:
\begin{equation}
    \log \gamma_t(\sigma,\sigma') = \log P(x_t(\sigma,\sigma')) + \log P(f_t(\sigma,\sigma')).
\end{equation}
Given that, by our definition of LLR, it is easy to see that:
\begin{equation}
   \begin{cases}
        P(x_t=0) = e^{L(x_t)}/(1+e^{L(x_t)})\\
        P(x_t=1) = 1/(1+e^{L(x_t)}),
    \end{cases}
\end{equation}
and the same holds for $f_t$. Thus, the logarithmic branch transition probabilities can be explicitly calculated as a function of the branch label:
\begin{equation}
\small
    \begin{cases}
        \log\gamma_t(0/0) = L(x_t) -\log(1+e^{L(x_t)}) + L(f_t) - \log (1+e^{L(f_t)})  \\
        \log \gamma_t(1/1) = -\log(1+e^{L(x_t)}) - \log (1+e^{L(f_t)}) \\ 
        \log \gamma_t(0/1) = -\log(1+e^{L(x_t)}) + L(f_t) - \log (1+e^{L(f_t)})  \\ 
        \log\gamma_t(1/0) =  L(x_t) -\log(1+e^{L(x_t)}) - \log (1+e^{L(f_t)}). \\ 
    \end{cases}
\end{equation}
The logarithmic forward and backward metrics are calculated recursively with the approximated formula:
\begin{equation}
\begin{cases}
      \log \alpha_t(\sigma) \approx \max_{\sigma'\in\Sigma}  \left[\log \alpha_{t-1}(\sigma) + \log\gamma_t(\sigma,\sigma') \right] \\
       \log \beta_t(\sigma) \approx \max_{\sigma'\in\Sigma} \left[\log\gamma_{t+1}(\sigma,\sigma') + \log \beta_{t+1}(\sigma)  \right].
\end{cases}
\label{eq:for_back}
\end{equation}
Finally, $L'(x_t)$ is calculated as:
\begin{equation}
    \begin{aligned}
           L'(x_t) \approx & \max_{(\sigma,\sigma')\in \mathcal T_0}\left[\log \alpha_{t-1}(\sigma) + \log\gamma_t(\sigma,\sigma') + \log \beta_t(\sigma') \right] - \\ 
     & \max_{(\sigma,\sigma')\in \mathcal T_1}\left[\log \alpha_{t-1}(\sigma) + \log\gamma_t(\sigma,\sigma') + \log \beta_t(\sigma') \right], 
    \end{aligned}
     \label{eq:out_bcjr}
\end{equation}
where $\mathcal T_0,\mathcal T_1$ are the sets of branches with output label equal to 0 and 1, respectively, at the time $t$. It is useful to introduce now the concept of \textit{extrinsic information}, which is the portion of output LLRs from the BCJR algorithm that does not depend on the input LLRs. Such extrinsic information is defined for $x_t$ as:
\begin{equation}
    L_{EXT}(x_t) = L'(x_t) - L(x_t).
    \label{eq:extrinsic}
\end{equation}


\section{Modeling of hook errors}
\label{sec:modeling}
In this Section, we first illustrate how to construct a \textit{joint Tanner graph} that takes into account errors on the data qubits as well as faults on the ancilla qubits. We get inspiration from the classical decoding problem of LDPC codes over the ISI channel \cite{siegel_isi2,siegel_isi}, where equalization and decoding are performed jointly on two interconnected Tanner graphs. To further optimize this procedure, we exploit the temporal correlation of hook errors happening on one ancilla qubit but at different time instants; in this way, we are able to remove short cycles from the joint Tanner graph, and to use trellis-based decoding for each individual ancilla.

\subsection{Construction of the joint Tanner graph}
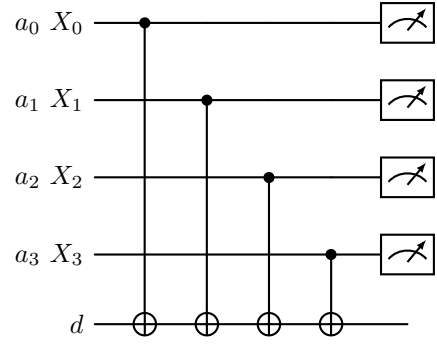
\begin{figure}
    \centering
\input{tikz/circuit2.tikz}
    \caption{Propagation of $X$ errors from $X$ stabilizer measurements on a single data qubit.}
    \label{fig:qubit_error}
\end{figure}

\begin{figure}
    \centering
    \input{tikz/decoding2.tikz}
    \caption{Example of the extended Tanner graph for a $(2,4)$-regular QLDPC code. The top layer of nodes are the codes' check nodes, represented as red squares; following, there are the code's variable nodes, represented as blue circles. To each variable node it is connected a constraint node, which can be interpreted as an always satisfied check node. The last layer consists of additional variable nodes corresponding to sources of hook errors.}
    \label{fig:extended_t1}
\end{figure}
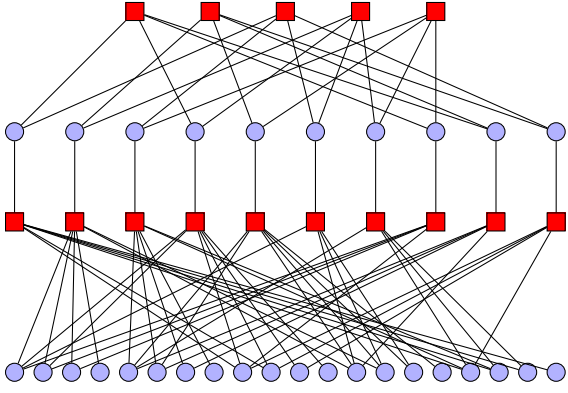

We proceed to construct the joint Tanner graph. By analyzing the circuit in Fig. \ref{fig:experiment}, we observe that $X$ errors on $X$ ancillas in the first slice propagate to the data qubits through $a_1, b_1, a_2,$ and $b_2$. Similarly, $X$ errors in the second slice propagate through $b_1, a_2,$ and $b_2$, and so on. It is easy to extend this reasoning for the case of stabilizer weight $\rho$. 
In Fig. \ref{fig:qubit_error} we illustrate an example of the interaction of a data qubit with several ancillas, where each CNOT gate is part of a different stabilizer measurement. It is evident that, from the data qubit perspective, the probability of error is equal to the probability of having an odd number of errors in the ancilla qubits before the CNOT interaction, and that errors originating from different ancilla qubits are independent. Namely, let $P_e(a_i)$ be the probability of having an $X$ error on the ancilla $a_i$ before the stabilizer measurement, then the probability of error on the data qubit $P_e(d)$ is:
\begin{equation}
    P_e(d) = \frac{1}{2}-\frac{1}{2}\prod_{i=1}^{\gamma}(1-2P_e(a_i)).
\end{equation}
It naturally follows that we can interpret each ancilla qubit $a_i$ as a bit, and the data qubit $d$ as a parity check equation. By doing so for every data qubit, we can construct the parity check matrix $\mathbf{P}$, where each row corresponds to a data qubit, each column represents an $X$ error on some ancilla qubit at a certain time-step, and nonzero entries indicate error propagation from the ancilla to the data qubits. For our experiment, we can conveniently express $\mathbf{P}$ in the polynomial form $P(x)$, illustrated in (\ref{eq:m}).

\begin{figure*}
  \begin{equation}
   P(x)= \left( \begin{array}{c|c|c|c|c|c|c}
        b(x) & b(x) & b(x)\setminus x^{b_1} & b(x)\setminus x^{b_1}  & b(x)\setminus (x^{b_1}+ x^{b_2}) & ...& b(x)\setminus (x^{b_1}+...+ x^{b_{d_v-1}}) \\ \hline
        a(x) & a(x)\setminus x^{a_1} & a(x)\setminus x^{a_1} & a(x)\setminus (x^{a_1}+x^{a_2}) & a(x)\setminus (x^{a_1}+x^{a_2}) & ... & 0
    \end{array}\right)
    \label{eq:m}
\end{equation}  
\end{figure*}

The joint Tanner graph is then constructed as follows: we take the original Tanner graph of $\mathbf{H}_Z$, namely $\mathcal T_Z$, and the Tanner graph related to the matrix $\mathbf{P}$, namely $\mathcal T_P$. We create the joint Tanner graph $\mathcal T_J$ by connecting each variable node of $\mathcal T_Z$ with one check node of $\mathcal T_P$. Such Tanner graph can be described by the matrix 
\begin{equation}
    \mathbf{H}_J = \begin{pmatrix}
        \mathbf{H}_Z & \mathbf{0} \\ 
        \mathbf{I}_n & \mathbf{P}
    \end{pmatrix}.
\end{equation}
We call the check nodes of $\mathcal T_P$ \textit{constraint nodes}, which are to be considered always satisfied. An example of $\mathcal{T}_J$ is illustrated in Fig.~\ref{fig:extended_t1}. A message-passing decoder operating on the joint Tanner graph is able to estimate errors occurring on data qubits, as well as faults originating hook errors. It follows that $\mathcal{T}_J$ constitutes an alternative to the standard circuit-level Tanner graph. Nevertheless, the graph $\mathcal{T}_J$ still retains most of the undesired properties of the circuit-level Tanner graph: indeed, a large amount of 4-cycles will always be present in $\mathbf{P}$, regardless of the code's original graph and of the measurement schedule. Moreover, the variable node degree in $\mathbf{P}$ is still irregular, with several variable nodes having degree 1. We notice that it is possible to exploit the correlation of hook errors arising from the same ancilla in different time steps to simplify $\mathcal T_J$, removing all the short cycles and the low-degree variable nodes.


\subsection{Single stabilizer measurement}
\label{sec:single_stab}

Consider the simple case of a hook error occurring on qubit $a$ in a single stabilizer measurement circuit, with stabilizer weight $\rho$. An example for $\rho=4$ is illustrated in Fig.~\ref{fig:hook_error}, where the qubit $a$ serves as an ancilla, while qubits $d_1, \dots, d_4$ are data qubits. 
We denote by $X_t$ an $X$-error affecting qubit $a$ at time step $t$. The propagation pattern of these errors is as follows: $X_1$ spreads to $d_1, d_2,...,d_{\rho}$; $X_2$ propagates to $d_2,..., d_{\rho}$, and so on, until $X_{\rho}$ propagates only to $d_{\rho}$.
To systematically represent this propagation, we define a binary matrix $\mathbf{G} \in \mathbb{F}_2^{\rho \times \rho}$, where each column corresponds to an error $X_t$, each row represents an error on the data qubit $d_j$, and an entry of 1 at position $(t, j)$ indicates that the error $X_t$ propagates to qubit $d_j$. It is straightforward to see that the matrix $\mathbf{G}$ has upper triangular form:

\begin{equation}
    \mathbf{G} = \begin{pmatrix}
        1 & 1 & \cdots & 1 \\
        1 & 1 & \cdots &0 \\
        1 & 1 & \iddots & \vdots\\
        1 & 0 & \cdots &0
    \end{pmatrix}.
        \label{eq:g}
\end{equation}
The structured form of $\mathbf{G}$ enables us to model the relationship between errors and their propagation effectively. Let $\mathbf{x}, \mathbf{d} \in \mathbb{F}_2^{\rho}$ be vectors satisfying $\mathbf{x} \cdot \mathbf{G}^T = \mathbf{d}$. From this, we obtain the following recursive relations:
\begin{equation}
  \begin{cases}
    d_1 = x_1 \\
    d_2 =  x_1 + x_2 = d_1 + x_2 \\ 
    \vdots \\
    d_t = d_{t-1} + x_t.
  \end{cases}
\label{eq:symbols}
\end{equation}
This naturally leads to the interpretation that the propagation of $X$ errors across time steps can be modeled as an error source with memory. Specifically, one can conceptualize an encoding circuit that represents $\mathbf{G}$, illustrated in Fig. \ref{fig:accumulator}.
In this circuit, information flows along the arrows, with the block $D$ acting as a memory element. The memory block outputs at time $t$ the input it received at time $t-1$, effectively storing state information. Consequently, the state of the memory element (and the entire circuit) at time $t$ corresponds to its input at time $t-1$, restricting the possible states to 0 and 1. The symbol $\oplus$ represents the XOR operation applied to its two inputs.

The circuit in Fig. \ref{fig:accumulator} can be interpreted as a finite-state machine (FSM), with its state diagram and corresponding trellis illustrated in Fig. \ref{fig:fsm} and Fig. \ref{fig:trellis}, respectively. 
We now compute the probability of error for each data qubit involved in the stabilizer measurement of Fig. \ref{fig:hook_error}. We showed previously that $d_t = d_{t-1}\oplus x_t$. Assuming that $p(x_t = 1) = p$ for every $t$, it easy to see that the probability of error for each qubit $d_t$ is equal to the probability of having an odd number of errors in $X_{t},X_{t-1},...,X_1$:
\begin{equation}
    P_e(d_t) = \frac{1-(1-2p)^{t}}{2}.
\end{equation}
Moreover, it is evident how the probability of error $P_e(d_t)$ and $P_e(d_{t-1})$ are correlated, such that:
\begin{equation}
    P_e(d_t) = p + P_e(d_{t-1}) - pP_e(d_{t-1}).
\end{equation}

\begin{figure}
    \centering
    \input{tikz/accumulator.tikz}
    \caption{The circuit representation of a hook error. Errors on the ancilla qubit $a$ at the time step $t$ are represented by the bit $x_t$, while $d_t$ represents the error on the corresponding data qubit. The memory element $D$ is a linear shift register.}
    \label{fig:accumulator}
\end{figure}
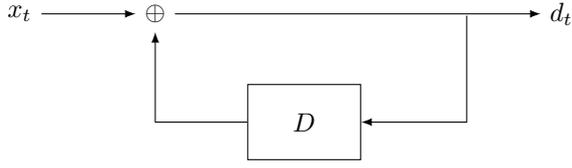

\begin{figure}
    \centering
    \input{tikz/fsm.tikz}
    \caption{Finite state machine corresponding to the circuit of Figure \ref{fig:accumulator}. Nodes correspond to states, edges represent state transitions. Edge labels have the form input/output, such that a particular input corresponds an output and to a state transition.}
    \label{fig:fsm}
\end{figure}
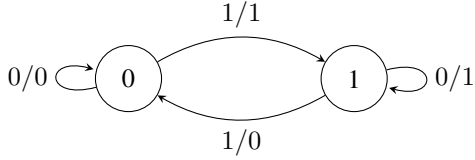

By modeling a hook error originating from a stabilizer measurement as an error source with memory and representing it with a trellis, we can leverage the BCJR algorithm to perform equalization. 
Notably, the leftmost portion of $P(x)$ coincides with $H_X(x)^T$, while all subsequent columns lie within the support of those in the first segment. Faults happening in the first slice of Fig. \ref{fig:experiment} correspond to columns in the first segment (starting from the left); faults happening in the second slice correspond to columns in the second segment, and so on.

Let us rearrange the columns of $\mathbf{P}$ as follows: we group together the $i$-th column of each segment, recognizing that the $i$-th column of the $j$-th segment is contained within the corresponding columns of all preceding segments ${(j-1), (j-2), \dots, 1}$. The resulting matrix is structured as:
\begin{equation}
    \mathbf{P}' = [\mathbf{P}^{(1)}\ \mathbf{P}^{(2)}\ ...\ \mathbf{{P}}^{(\gamma)}].
\end{equation}

Let the support of a vector $\supp(\mathbf{v}) = \{j\ |\ v_j\neq 0\}$ be the set of indices of its non-zero entries; let $\mathbf{p}^{(k)}_{i,:}$ denote the $i$-th row of $\mathbf{P}^{(k)}$, and $\mathbf{p}^{(k)}_{:,j}$ denote the $j$-th column of $\mathbf{P}^{(k)}$. It is clear that $\supp(\mathbf p^{(k)}_{:,1}) = \supp (\mathbf{H}_{X_{:,k}}^T)$, and that $\supp(\mathbf p^{(k)}_{:,j}) \subset \supp(\mathbf p^{(k)}_{:,1})$, for all $j=2,..,d_c$.

\begin{lemma}
\label{lemma}
    Let $K = \supp (\mathbf{H}_{X_{:,k}}^T)$ be the support of the $k$-th column of $\mathbf{H}_X^T$, and let $\mathbf P^{(k)}_{K,:}$ be the restriction of the rows of  $\mathbf P^{(k)}$ to the indices in $K$. Then, there always exist a row permutation $\pi$ such that $\pi\mathbf P^{(k)}_{K,:} = \mathbf{G}$.
    \begin{proof}
        We can divide $\mathbf P^{(k)}_{K,:}$ in two blocks, just as we did for $P(x)$. Each monomial of $a(x)$ and $b(x)$ corresponds to a 1 $\mathbf p^{(k)}_{K,1}$, which is an all-one column, as the first segment of $P(x)$ contains both the full polynomials. The second column $\mathbf p^{(k)}_{K,2}$ is equal to the first column, with the omission of the monomial $x^{a_1}$, hence, it has weight $\rho - 1$ and its support is strictly contained in that of the first. Continuing this process, the $j$-th column $\mathbf{p}^{(k)}_{K,j}$ omits $j - 1$ specific monomials and therefore has weight $\rho  - (j - 1)$. Thus, $\mathbf{P}^{(k)}_{K,:}$ has columns with strictly decreasing weights, where the support of each column is nested within that of all previous columns.  The same nested structure also holds for the rows of $\mathbf{P}^{(k)}_{K,:}$. Next, we sort the rows in order of increasing Hamming weight. Since the leftmost column is all ones, the row with weight 1 must have its sole nonzero entry in the first column and thus appears at the top. The row with weight 2 has ones in the first two columns, and so on, until we reach the all-one row, which has weight $\rho$ and is placed at the bottom. As a result, the nonzero entries in each row are aligned to the left and form a staircase pattern down the matrix, yielding a lower triangular structure.
    \end{proof}
\end{lemma}
Lemma \ref{lemma} shows that there exists a permutation $\pi_i$ to reduce the restriction of $\mathbf{P}^{(i)}$ on its nonzero rows to the matrix $\mathbf{G}$. We illustrate this procedure with the following example.

\begin{figure}
    \centering
    \input{tikz/trellis.tikz}
    \caption{Trellis for the finite state machine of Figure \ref{fig:fsm}, assuming a code with $d_c=4$. Nodes and edges correspond to the one of a FSM, with every layer representing the possible transitions at the time step $t$. }
    \label{fig:trellis}
\end{figure}
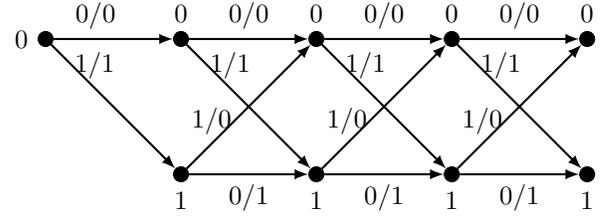

\begin{example}
    Consider the following $(2,4)$-regular bicycle code with $N=5$:
\begin{align*}
    \mathbf{H}_z = &
\left[ \begin{array}{ccccc|ccccc}
1 & 0 & 0 & 1 & 0 & 0 & 0 & 1 & 0 & 1 \\
0 & 1 & 0 & 0 & 1 & 1 & 0 & 0 & 1 & 0 \\
1 & 0 & 1 & 0 & 0 & 0 & 1 & 0 & 0 & 1 \\
0 & 1 & 0 & 1 & 0 & 1 & 0 & 1 & 0 & 0 \\
0 & 0 & 1 & 0 & 1 & 0 & 1 & 0 & 1 & 0
\end{array}\right]\\
\mathbf{H}_x = &
\left[ \begin{array}{ccccc|ccccc}
0 & 1 & 0 & 1 & 0 & 1 & 0 & 1 & 0 & 0 \\
0 & 0 & 1 & 0 & 1 & 0 & 1 & 0 & 1 & 0 \\
1 & 0 & 0 & 1 & 0 & 0 & 0 & 1 & 0 & 1 \\
0 & 1 & 0 & 0 & 1 & 1 & 0 & 0 & 1 & 0 \\
1 & 0 & 1 & 0 & 0 & 0 & 1 & 0 & 0 & 1
\end{array}\right]
\end{align*}
The corresponding polynomial representation is
   \begin{align*}
    H_Z(x) = & [1 + x^3, x^2 + x^4] \\
    H_X(x) = & [x + x^3, 1 + x^2]
\end{align*} 

The matrix $\mathbf P$, given the measurement circuit in Fig. \ref{fig:experiment}, is illustrated in  (\ref{eq:example}).

Let us consider the first column of each block, and construct $\mathbf{P}^{(1)}$, that has the form:
\begin{equation*}
    \mathbf{P}^{(1)} = \left(\begin{array}{cccc}
        0 & 0 & 0 & 0 \\
         1 & 1 & 1 & 1 \\
         0 & 0 & 0 & 0 \\
         1 & 1 & 0 & 0 \\
         0 & 0 & 0 & 0 \\
         1 & 0 & 0 & 0 \\
         0 & 0 & 0 & 0 \\
         1 & 1 & 1 & 0 \\
         0 & 0 & 0 & 0 \\
         0 & 0 & 0 & 0
    \end{array}\right)
\end{equation*}
By removing the all-zero rows, and by properly permuting rows it is easy to see that we get:
\begin{equation*}
     \pi_1(\mathbf{P}^{(1)}_{K,:}) =   \mathbf{G} = \begin{pmatrix}
        1 & 1 & 1 & 1 \\
        1 & 1 & 1 &0 \\
        1 & 1 & 0 & 0\\
        1 & 0 & 0 &0
    \end{pmatrix}.
\end{equation*}
We can apply the same analysis systematically for all the columns of the first block, and the result does not change (besides the permutation $\pi_i$).
\end{example}

\begin{figure*}
    \begin{equation}
        \mathbf{P} =
\left( \begin{array}{ccccc|ccccc|ccccc|ccccc}
0 & 0 & 1 & 0 & 1 & 0 & 0 & 1 & 0 & 1 & 0 & 0 & 0 & 0 & 1 & 0 & 0 & 0 & 0 & 1 \\
1 & 0 & 0 & 1 & 0 & 1 & 0 & 0 & 1 & 0 & 1 & 0 & 0 & 0 & 0 & 1 & 0 & 0 & 0 & 0\\
0 & 1 & 0 & 0 & 1 & 0 & 1 & 0 & 0 & 1 & 0 & 1 & 0 & 0 & 0 & 0 & 1 & 0 & 0 & 0\\
1 & 0 & 1 & 0 & 0 & 1 & 0 & 1 & 0 & 0 & 0 & 0 & 1 & 0 & 0 & 0 & 0 & 1 & 0 & 0\\
0 & 1 & 0 & 1 & 0 & 0 & 1 & 0 & 1 & 0 & 0 & 0 & 0 & 1 & 0 & 0 & 0 & 0 & 1 & 0\\ \hline
1 & 0 & 0 & 1 & 0 & 0 & 0 & 0 & 1 & 0 & 0 & 0 & 0 & 1 & 0 & 0 & 0 & 0 & 0 & 0\\
0 & 1 & 0 & 0 & 1 & 0 & 0 & 0 & 0 & 1 & 0 & 0 & 0 & 0 & 1 & 0 & 0 & 0 & 0 & 0\\
1 & 0 & 1 & 0 & 0 & 1 & 0 & 0 & 0 & 0 & 1 & 0 & 0 & 0 & 0 & 0 & 0 & 0 & 0 & 0 \\
0 & 1 & 0 & 1 & 0 & 0 & 1 & 0 & 0 & 0 & 0 & 1 & 0 & 0 & 0 & 0 & 0 & 0 & 0 & 0 \\
0 & 0 & 1 & 0 & 1 & 0 & 0 & 1 & 0 & 0 & 0 & 0 & 1 & 0 & 0 & 0 & 0 & 0 & 0 & 0
\end{array} \right)
\label{eq:example}
    \end{equation}
\end{figure*}
This procedure can be applied to all submatrices $\mathbf{P}^{(i)}$, and since each one of them is associated with a trellis, we merge all variable nodes corresponding to columns of $\mathbf{P}^{(i)}$, creating \textit{equalizer nodes}. 
Readers familiar with doubly generalized LDPC codes \cite{dgldpc} will recognize that equalizer nodes function as generalized variable nodes. Each equalizer node represents the $\rho$ faults from a single ancilla qubit and produces $\rho$ output messages corresponding to the outputs of the SISO estimator. Consequently, $\mathcal T_P$ reduces to $\mathcal T_X^T$, which is the Tanner graph of $\mathbf{H}_X^T$. Hence, the joint Tanner graph reduces to:
\begin{equation}
    \mathbf{H}_J = \begin{pmatrix}
        \mathbf{H}_Z & \mathbf{0} \\ 
        \mathbf{I}_n & \mathbf{H}_X^T
    \end{pmatrix}.
\end{equation}
An illustration of $\mathcal T_J$ is illustrated in Fig.~\ref{fig:extended_t}.
The resulting scenario is as follows: each $X$ stabilizer measurement acts as an independent error source with memory, similarly to a classical ISI channel. This similarity allows us to employ trellis-based equalization techniques for decoding. However, since each of these independent noise sources affects different subsets of qubits, our turbo equalizer architecture requires a dedicated trellis equalizer for each $X$ stabilizer. As we will demonstrate in the following Section, these equalizers not only exchange soft information with the BP decoder but also interact with one another to refine the overall error correction process.
Although we illustrate this process for BB codes, we remark that an analogous process is possible for any family of QLDPC codes.

\section{Turbo annihilation decoder}
\label{sec:decoder}
In this section, we describe the decoder architecture designed to handle correlated noise induced by hook errors in quantum circuits. Our approach extends the well-known min-sum (MS) \cite{nms_decoder} algorithm by incorporating side information on error correlations among data qubits. This is achieved through a series of soft-input soft-output (SISO) estimators operating on a trellis. In this paper we utilize BCJR estimators, however there exists a variety of different approaches, such as the \textit{minimum mean square error} (MMSE) estimation \cite{turbo_eq}. The decoder itself operates on the joint Tanner graph illustrated in Fig.~\ref{fig:extended_t}.

At each iteration, messages are exchanged between variable nodes, equalizer nodes, and check nodes. We define $\nu_{j,i}^{\ell}$ as the variable-to-check message from variable node $j$ to check node $i$ at iteration $\ell$, $\sigma^{\ell}_{j,i}$ as the equalizer-to-check message from equalizer node $j$ to check node $i$ at iteration $\ell$, and $\mu_{i,j}^{\ell}$ as the check-to-variable message from check node $i$ to variable node $j$ at iteration $\ell$.
The check-to-variable update follows the standard MS rule:
\begin{equation}
    \mu_{i,j}^{\ell} = (1-2s_i) \prod_{j' \in \mathcal{N}(i) \setminus j} \mathrm{sgn}(\nu_{j',i}^{\ell-1}) \cdot \min |\nu_{j',i}^{\ell-1}|,
\end{equation}
where $\mathcal{N}(i)$ denotes the set of neighbors of check node $i$. The sign function is defined as:
\begin{align}
    \mathrm{sgn}(x) = \begin{cases}
        -1, & x < 0, \\
        +1, & \text{otherwise}.
    \end{cases}
\end{align}
For constraint nodes, we set $s_i = 0$. 
Variable node updates differ between standard variable nodes and equalizer nodes. The variable-to-check message is computed as:
\begin{equation}
    \nu_{j,i}^{\ell} = \sum_{i' \in \mathcal{N}(j) \setminus i} \mu_{i',j}^{\ell}.
    \label{eq:v2c}
\end{equation}
Unlike standard MS decoding, there are no explicit prior channel LLRs associated with the variable nodes. Instead, for the original Tanner graph, the prior information comes from the constraint node messages, while for the equalizer graph, the prior information is the sum of the incoming check node messages.
Thus, prior LLRs are already incorporated into Eq.~\eqref{eq:v2c}.

Each SISO estimator exchanges extrinsic information with the constraint nodes. The input messages to the $j$-th equalizer node, $\mu^{\ell}_{1,j}, \dots, \mu^{\ell}_{\rho,j}$, are first permuted according to $\pi_j$, the permutation used to transform $\mathbf{M}_j$ into $\mathbf{G}$, yielding $\mu^{\ell}_{\pi_j(1),j}, \dots, \mu^{\ell}_{\pi_j(\rho),j}$. 
Additionally, the SISO estimator receives prior LLRs for each fault $L_j(f_1), \dots, L_j(f_{\rho})$. Based on this information, it computes a set of \textit{a posteriori} LLRs, $L_{j,\pi_j(1)}^{\ell}, \dots, L_{j,\pi_j(\rho)}^{\ell}$. Applying $\pi_j^{-1}$ to the output, the message sent to the $i$-th constraint node corresponds to the extrinsic information:
\begin{equation}
    \sigma^{\ell}_{j,i} = L_{j,\pi_j^{-1}(i)}^{\ell} - \mu_{i,j}^{\ell}.
\end{equation}
At each iteration, the \textit{a posteriori} LLR for variable node $j$ is computed as:
\begin{equation}
    q_j^{\ell} = \sum_{i \in \mathcal{N}(j)} \mu_{i,j}^{\ell},
\end{equation}
and the hard decision is made based on:
\begin{equation}
    \hat{e}_j^{\ell} = 
    \begin{cases}
        0, & q_j^{\ell} \geq 0, \\
        1, & q_j^{\ell} < 0.
    \end{cases}
\end{equation}
The decoder runs until the estimated error vector $\hat{\mathbf{e}}$ satisfies the measured syndrome or until a predefined maximum number of iterations is reached.

\subsection{Complexity analysis}
We analyze the complexity for a $(\gamma, \rho)$-regular QLDPC code with $n$ variable nodes and $m$ check nodes. In each iteration each variable node sends $\gamma+1$ messages, and each check node sends $\rho$ messages, thus, the decoder performs $N_O = n(\gamma+1) + m \rho$ operations on the original Tanner graph.
Each equalizer node runs the BCJR algorithm, whose complexity depends on the trellis size. The trellis has two states, each one with two incoming edges, and $\rho$ layers. Each layer requires 4 operations for forward/backward metric updates and 5 additional operations to compute output LLRs.
Thus, the BCJR complexity per layer is $N^L_{\text{BCJR}} = 9$ operations, leading to a total of $N_{\text{BCJR}} = 9 \rho$ per equalizer. Each constraint node computes $\gamma+1$ check-to-variable messages, resulting in:
\begin{equation}
    N_E = 9m \rho + n(\gamma+1).
\end{equation}
The total complexity per iteration is:
\begin{equation}
    N = 2n(\gamma+1) + 10m \rho,
    \label{eq:complexity}
\end{equation}
which scales as $\mathcal{O}(n)$.

\begin{figure}
    \centering
    \input{tikz/decoding.tikz}
    \caption{Example of the extended Tanner graph for a $(2,4)$-regular QLDPC code. The top layer of nodes are the codes' check nodes, represented as red squares; following, there are the code's variable nodes, represented as blue circles. To each variable node it is connected a constraint node, which can be interpreted as an always satisfied check node. The last layer consists of equalizer nodes, represented as large blue circles, where ad hoc permutations to the edges are applied.}
    \label{fig:extended_t}
\end{figure}

\subsection{Convergence improvement}
During our decoder experiments, we encountered several challenges related to convergence. The combination of error patterns and the presence of trapping sets in the decoding graph often prevents the decoder from converging to a valid error estimate. Since we aim to avoid using Ordered Statistics Decoding (OSD), we adopt several heuristics to enhance convergence while maintaining linear complexity in the blocklength.
Our primary strategy is the already mentioned decoding diversity.
In our implementation, we use three decoders. The first two employ a \textit{layered schedule}, which we describe below, while the third uses a flooding (or parallel) schedule.

The layered schedule reflects the natural flow of information from each fault through the decoding graph, mirroring the structure of the circuit. At each iteration, messages propagate sequentially: first from the equalizers to the adjacent constraint nodes, then to the variable nodes. From there, messages are passed to the check nodes and then returned to the variables. The updated variable messages are subsequently sent back to the constraint nodes, which in turn update the equalizers. Finally, a hard decision is made on the variable nodes based on the current messages.

In all three decoders, we modify the variable-to-check update rule using the \textit{Min-Sum with Past Influence (MS-PI)} algorithm~\cite{chytas2025enhancedminsumdecodingquantum}, which is specifically designed for BB codes. In MS-PI, the update for a variable node labeled \( L \) or \( R \) is defined as:
\begin{equation}
\nu^{(\ell)}_{j,i} = 
\begin{cases}
\nu, & \text{if } \mathrm{sgn}(\nu) = \mathrm{sgn}(\nu^{(\ell-1)}_{j,i}) \\
\nu + \nu^{(\ell-1)}_{j,i}, & \text{otherwise},
\end{cases}
\label{eq:Pi}
\end{equation}
where \( \nu^{(\ell)}_{j,i} \) denotes the message passed from variable node \( j \) to check node \( i \) at iteration \( \ell \), and \( \nu \) is the current message computed from neighboring check nodes.
MS-PI has been shown to closely approach the performance of BPOSD0 when decoding BB codes over depolarizing noise. In our setting, we also observe a notable performance improvement. Specifically, the first and third decoders apply the MS-PI rule~\eqref{eq:Pi} to variable nodes labeled \( L \), while the second decoder applies it to those labeled \( R \).
Additionally, all decoders use a normalization factor of \( \beta = 0.875 \) in the check update rule to help stabilize message magnitudes and improve convergence.


\section{Simulation results}
\label{sec:results}
We assess the performance of our decoder with numerical simulations. We consider the scenario of decoding the $\llbracket 90,8,10 \rrbracket$ and the $\llbracket 144,12,12 \rrbracket$ BB codes from \cite{bravyi2024high}, with a measuring setting equivalent to the one in Fig.~\ref{fig:experiment} (except for the fact that BB codes is $(3,6)$-regular). 
We use the $\mathtt{stim}$ package \cite{gidney2021stim} to sample the errors from our circuit. We compare the turbo annihilation (TA) decoder with the standard normalized MS with normalization parameter $\beta=0.875$ and the BPOSD0 decoder, both running on the circuit-level Tanner graph. We utilize the diversity decoding strategy as described before, employing a total of 3 decoders, each one running for a maximum of 300 iterations. Similarly, the BPOSD0 runs for a maximum of 300 iterations before the post-processing phase. The MS decoder, in contrast, runs for 900 iterations, to fairly compare it with TA. For each value of $p$ being the fault probability for each noisy component in the circuit, \textit{i.e.}, the probability of having depolarizing noise on data qubits and $X$ ancillas at the beginning of the circuit and the probability of CNOT faults, we evaluate the logical error rate as the ratio between the number of logical errors and the number of trials. The minimum number of failures to stop the simulation is set to 10, while the minimum amount of trials is set to $10^7$. 

\begin{figure}
    \centering
    \includegraphics[width=1\linewidth]{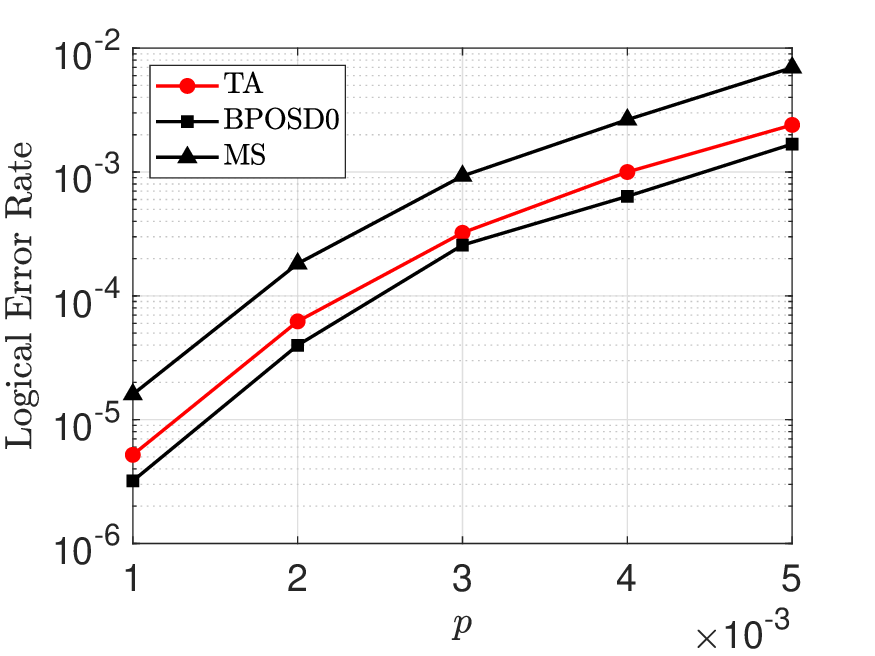}
    \caption{Performance comparison between TA, MS and BPOSD0, for the $\llbracket 90,8,10 \rrbracket$ BB code.}
    \label{fig:simulation90}
\end{figure}

In Fig.~\ref{fig:simulation90} we illustrate the results for the $\llbracket 90,8,10 \rrbracket$ code. The abscissa axis represents fault probability $p$ in the circuit, while the ordinate axis represents the probability of logical error. It is evident how our decoder performs consistently better than MS, closely approaching BPOSD0 for all the values of $p$. \

\begin{figure}
    \centering
    \includegraphics[width=1\linewidth]{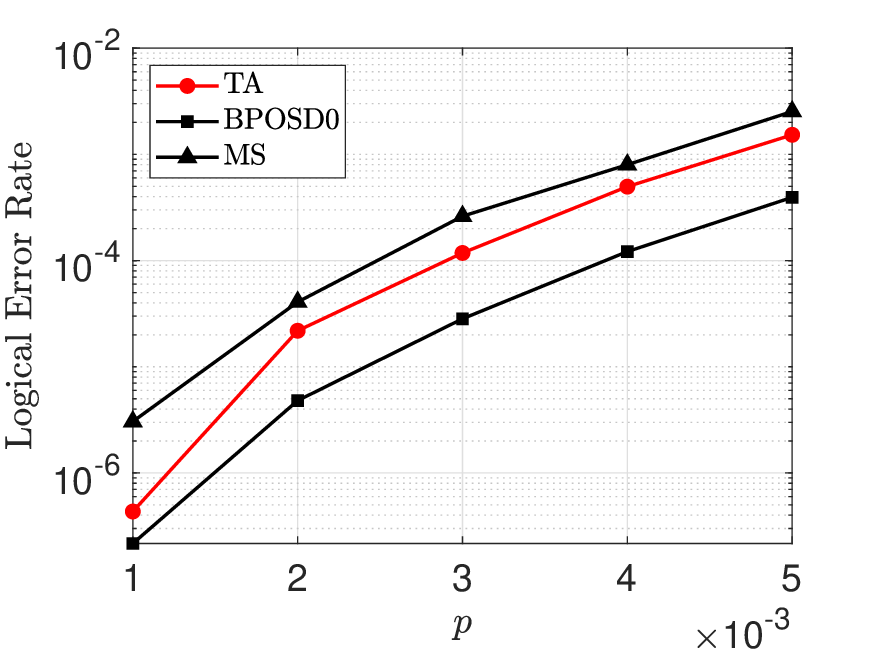}
    \caption{Performance comparison between TA, MS and BPOSD0, for the $\llbracket 144,12,12 \rrbracket$ BB code.}
    \label{fig:simulation144}
\end{figure}

In Fig.~\ref{fig:simulation144} we illustrate the results for the $\llbracket 144,12,12 \rrbracket$ code (also known as \textit{Gross} code). We highlight how the TA decoder still performs better than MS, although with a smaller gap than the previous case; nevertheless, it is able to approach BPOSD0 performance for low enough probability of error.

\section{Conclusions}
In this paper, we introduced a new decoding framework for correcting correlated noise in quantum circuits. Unlike the traditional circuit-level Tanner graph approach, which tries to determine the exact location of faults, our method focuses on estimating the overall effect of those faults on the data qubits. We take advantage of the time correlation between hook errors coming from the same ancilla qubits to reduce the size of the decoding graph and eliminate short cycles and low-degree variable nodes. Our decoder combines a set of trellis-based BCJR equalizers with a standard min-sum decoder to perform turbo equalization of the correlated noise. This approach achieves significantly better performance than MS on the circuit-level Tanner graph, and it is able to come close to the performance of BPOSD0 while maintaining $\mathcal{O}(n)$ complexity. Because it operates on a much smaller graph that preserves the structure of the original code, the decoder is scalable and well-suited for large quantum systems. Future work includes implementing the TA decoder in standard quantum memory experiments and systematically evaluating its performance in scenarios with more complex noise correlations. Additionally, exploring more efficient SISO estimators could eliminate the need for BCJR, leading to further reductions in decoding complexity.

\bibliographystyle{ieeetr}

\input{referencesNEW}
\end{document}

%% file: tikz/circuit1.tikz
\begin{quantikz}
\lstick{$a$}\slice{$X_1$} &   \ctrl{1}\slice{$X_2$} & \ctrl{2}\slice{$X_3$} & \ctrl{3} \slice{$X_4$}& \ctrl{4} & \meter{} \\
\lstick{$d_1$}  & \targ{} &  &  & &  \\
\lstick{$d_2$}  &  &\targ{} &  &  & \\
\lstick{$d_3$}  &  &  & \targ{} &  & \\
\lstick{$d_4$} &  &  &  & \targ{} & 
\end{quantikz}

%% file: tikz/experiment.tikz
\begin{quantikz}
\lstick{$X$} \slice{}  & \ctrl[wire style={"a_1"}]{1} \gategroup[3,steps=4,style={dashed,rounded corners,fill=red!20, inner xsep=2pt},background,label style={label position=above,anchor=north,yshift=0.2cm}]{{\sc faulty}} \slice{} & \ctrl[wire style={"b_1", pos=0.8}]{2} \slice{} & \ctrl[wire style={"a_2"}]{1}  \slice{} & \ctrl[wire style={"b_2", pos=0.8}]{2} &  &  &  &  &  \\
\lstick{$L$}  & \targ{} &  & \targ{} &  & \ctrl{2} \gategroup[3,steps=4,style={dashed,rounded corners,fill=blue!20, inner xsep=1pt},background,label style={label position=below,anchor=north,yshift=-0.2cm}]{{\sc perfect}} &  & \ctrl{2} &  &  \\
\lstick{$R$}  &  & \targ{} &  & \targ{} &  & \ctrl{1} &  & \ctrl{1} &  \\
\lstick{$Z$}  &  & & &  &  \targ{} & \targ{} & \targ{} & \targ{}  & \meter{}
\end{quantikz}

%% file: tikz/circuit2.tikz
\begin{quantikz}
\lstick{$a_0\ X_0$} &   \ctrl{4} &  &  & & \meter{}\\
\lstick{$a_1\ X_1$}  &  & \ctrl{3} &  & & \meter{} \\
\lstick{$a_2\ X_2$}  &  & & \ctrl{2} &  & \meter{}\\
\lstick{$a_3\ X_3$}  &  &  & & \ctrl{1} &  \meter{}\\
\lstick{$d$} & \targ{} & \targ{} & \targ{} & \targ{} & 
\end{quantikz}

%% file: tikz/decoding2.tikz
\begin{tikzpicture}[scale=0.8, every node/.style={transform shape}]

\tikzstyle{circle_node} = [circle, draw, fill=blue!30, inner sep=0pt, minimum size=3mm]
\tikzstyle{square_node} = [rectangle, draw, fill=red, inner sep=0pt, minimum size=3mm]
\tikzstyle{perm_node} = [rectangle, minimum width=8.5mm, minimum height=5mm, draw, inner sep=0pt]
\tikzstyle{circle_alt_node} = [circle, draw, fill=blue!40, inner sep=0pt, minimum size=8.5mm]

\node[square_node] (c1) at (1, 0) {};
\node[square_node] (c2) at (2.25, 0) {};
\node[square_node] (c3) at (3.5, 0) {};
\node[square_node] (c4) at (4.75, 0) {};
\node[square_node] (c5) at (6, 0) {};

\node[circle_node] (b1) at (-1, -2) {};
\node[circle_node] (b2) at (0, -2) {};
\node[circle_node] (b3) at (1, -2) {};
\node[circle_node] (b4) at (2, -2) {};
\node[circle_node] (b5) at (3, -2) {};
\node[circle_node] (b6) at (4, -2) {};
\node[circle_node] (b7) at (5, -2) {};
\node[circle_node] (b8) at (6, -2) {};
\node[circle_node] (b9) at (7, -2) {};
\node[circle_node] (b10) at (8, -2) {};

\node[square_node] (q1) at (-1, -3.5) {};
\node[square_node] (q2) at (0, -3.5) {};
\node[square_node] (q3) at (1, -3.5) {};
\node[square_node] (q4) at (2, -3.5) {};
\node[square_node] (q5) at (3, -3.5) {};
\node[square_node] (q6) at (4, -3.5) {};
\node[square_node] (q7) at (5, -3.5) {};
\node[square_node] (q8) at (6, -3.5) {};
\node[square_node] (q9) at (7, -3.5) {};
\node[square_node] (q10) at (8, -3.5) {};

\node[circle_node] (e11) at (-1, -6) {};
\node[circle_node] (e12) at (-0.5263, -6) {};
\node[circle_node] (e13) at (-0.0526, -6) {};
\node[circle_node] (e14) at (0.4211, -6) {};
\node[circle_node] (e21) at (0.8947 , -6) {};
\node[circle_node] (e22) at (1.3684, -6) {};
\node[circle_node] (e23) at (1.8421, -6) {};
\node[circle_node] (e24) at (2.3158, -6) {};
\node[circle_node] (e31) at (2.7895, -6) {};
\node[circle_node] (e32) at (3.2632, -6) {};
\node[circle_node] (e33) at (3.7368, -6) {};
\node[circle_node] (e34) at (4.2105, -6) {};
\node[circle_node] (e41) at (4.6842, -6) {};
\node[circle_node] (e42) at (5.1579 , -6) {};
\node[circle_node] (e43) at (5.6316 , -6) {};
\node[circle_node] (e44) at (6.1053 , -6) {};
\node[circle_node] (e51) at (6.5789, -6) {};
\node[circle_node] (e52) at (7.0526, -6) {};
\node[circle_node] (e53) at (7.5263, -6) {};
\node[circle_node] (e54) at (8.0000, -6) {};

\draw(b1)--(q1);
\draw(b2)--(q2);
\draw(b3)--(q3);
\draw(b4)--(q4);
\draw(b5)--(q5);
\draw(b6)--(q6);
\draw(b7)--(q7);
\draw(b8)--(q8);
\draw(b9)--(q9);
\draw(b10)--(q10);

\draw(b1)--(c1);
\draw(b1)--(c3);
\draw(b2)--(c2);
\draw(b2)--(c4);
\draw(b3)--(c3);
\draw(b3)--(c5);
\draw(b4)--(c4);
\draw(b4)--(c1);
\draw(b5)--(c5);
\draw(b5)--(c2);
\draw(b6)--(c3);
\draw(b6)--(c4);
\draw(b7)--(c4);
\draw(b7)--(c5);
\draw(b8)--(c5);
\draw(b8)--(c1);
\draw(b9)--(c1);
\draw(b9)--(c2);
\draw(b10)--(c2);
\draw(b10)--(c3);

\draw(q2)--(e11);
\draw(q4)--(e11);
\draw(q6)--(e11);
\draw(q8)--(e11);
\draw(q2)--(e12);
\draw(q4)--(e12);
\draw(q8)--(e12);
\draw(q2)--(e13);
\draw(q8)--(e13);
\draw(q2)--(e14);

\draw(q3)--(e21);
\draw(q5)--(e21);
\draw(q7)--(e21);
\draw(q9)--(e21);
\draw(q3)--(e22);
\draw(q5)--(e22);
\draw(q9)--(e22);
\draw(q3)--(e23);
\draw(q9)--(e23);
\draw(q3)--(e24);

\draw(q1)--(e31);
\draw(q4)--(e31);
\draw(q8)--(e31);
\draw(q10)--(e31);
\draw(q1)--(e32);
\draw(q4)--(e32);
\draw(q10)--(e32);
\draw(q4)--(e33);
\draw(q10)--(e33);
\draw(q4)--(e34);

\draw(q2)--(e41);
\draw(q5)--(e41);
\draw(q6)--(e41);
\draw(q9)--(e41);
\draw(q2)--(e42);
\draw(q5)--(e42);
\draw(q6)--(e42);
\draw(q5)--(e43);
\draw(q6)--(e43);
\draw(q5)--(e44);

\draw(q1)--(e51);
\draw(q3)--(e51);
\draw(q7)--(e51);
\draw(q10)--(e51);
\draw(q1)--(e52);
\draw(q3)--(e52);
\draw(q7)--(e52);
\draw(q1)--(e53);
\draw(q7)--(e53);
\draw(q1)--(e54);

\end{tikzpicture}

%% file: tikz/accumulator.tikz
\begin{tikzpicture}[scale=1.8]

\node (x) at (0,0) {$x_t$};
\node (s) at (1,0) {$\oplus$};
\node (o) at (4,0) {$d_t$};
\node (m) [draw, rectangle, minimum width=1.5cm, minimum height=1cm] at (2.1,-.8) {$D$};
\node[inner sep=0pt] (i1) at (3.3,0) {};
\node (i2) at (3,-1) {};

\draw[-latex] (x)--(s);
\draw[-latex] (s)--(o);
\draw[-latex] (i1)|-(m);
\draw[-latex] (m)-|(s);
\end{tikzpicture}

%% file: tikz/fsm.tikz
\begin{tikzpicture}[->, >=stealth, node distance=3cm]
    \node[state] (S0) {0};
    \node[state, right of=S0] (S1) {1};

    \draw[->] (S0) edge[bend left] node[above] {$1/1$} (S1);
    \draw[->] (S1) edge[bend left] node[below] {$1/0$} (S0);
    \draw[->] (S0) edge[loop left] node[left] {$0/0$} (S0);
    \draw[->] (S1) edge[loop right] node[right] {$0/1$} (S0);
\end{tikzpicture}

%% file: tikz/trellis.tikz
\begin{tikzpicture}[scale=0.9]

\node (0) [draw, circle,inner sep=2pt, fill, black, label=left:$0$] at (-2,0) {};
\node (00) [draw, circle,inner sep=2pt, fill, black, label=above:$0$] at (0,0) {};
\node (01) [draw, circle,inner sep=2pt, fill, black, label=below:$1$] at (0,-2) {};
\node (10) [draw, circle,inner sep=2pt, fill, black, label=above:$0$] at (2,0) {};
\node (11) [draw, circle,inner sep=2pt, fill, black, label=below:$1$] at (2,-2) {};
\node (102) [draw, circle,inner sep=2pt, fill, black, label=above:$0$] at (4,0) {};
\node (112) [draw, circle,inner sep=2pt, fill, black, label=below:$1$] at (4,-2) {};
\node (103) [draw, circle,inner sep=2pt, fill, black, label=above:$0$] at (6,0) {};
\node (113) [draw, circle,inner sep=2pt, fill, black, label=below:$1$] at (6,-2) {};

\draw[-latex, thick] (0)--(00) node[above, pos=0.35] {$0/0$} ;
\draw[-latex, thick] (0)--(01) node[above, pos=0.35] {$1/1$} ;
\draw[-latex, thick] (00)--(11) node[above, pos=0.35] {$1/1$} ;
\draw[-latex,thick] (00)--(10) node[above, midway] {$0/0$};
\draw[-latex,thick] (01)--(11) node[below, midway] {$0/1$}  ;
\draw[-latex,thick] (01)--(10) node[above, pos=0.2] {$1/0$} ;
\draw[-latex, thick] (10)--(102) node[above, midway] {$0/0$} ;
\draw[-latex,thick] (10)--(112) node[above,  pos=0.35] {$1/1$};
\draw[-latex,thick] (11)--(112) node[below, midway] {$0/1$}  ;
\draw[-latex,thick] (11)--(102) node[above, pos=0.2] {$1/0$} ;
\draw[-latex, thick] (102)--(103) node[above, midway] {$0/0$} ;
\draw[-latex,thick] (102)--(113) node[above,  pos=0.35] {$1/1$};
\draw[-latex,thick] (112)--(113) node[below, midway] {$0/1$}  ;
\draw[-latex,thick] (112)--(103) node[above, pos=0.2] {$1/0$} ;
\end{tikzpicture}

%% file: tikz/decoding.tikz
\begin{tikzpicture}[scale=0.8, every node/.style={transform shape}]

\tikzstyle{circle_node} = [circle, draw, fill=blue!30, inner sep=0pt, minimum size=3mm]
\tikzstyle{circle_node2} = [circle, draw, fill=blue!30, inner sep=0pt, minimum size=2mm]
\tikzstyle{square_node} = [rectangle, draw, fill=red, inner sep=0pt, minimum size=3mm]
\tikzstyle{perm_node} = [rectangle, minimum width=8.5mm, minimum height=5mm, draw, inner sep=0pt]
\tikzstyle{circle_alt_node} = [circle, draw, fill=blue!40, inner sep=0pt, minimum size=8.5mm]

\node[square_node] (c1) at (1, 0) {};
\node[square_node] (c2) at (2.25, 0) {};
\node[square_node] (c3) at (3.5, 0) {};
\node[square_node] (c4) at (4.75, 0) {};
\node[square_node] (c5) at (6, 0) {};

\node[circle_node] (b1) at (-1, -2) {};
\node[circle_node] (b2) at (0, -2) {};
\node[circle_node] (b3) at (1, -2) {};
\node[circle_node] (b4) at (2, -2) {};
\node[circle_node] (b5) at (3, -2) {};
\node[circle_node] (b6) at (4, -2) {};
\node[circle_node] (b7) at (5, -2) {};
\node[circle_node] (b8) at (6, -2) {};
\node[circle_node] (b9) at (7, -2) {};
\node[circle_node] (b10) at (8, -2) {};

\node[square_node] (q1) at (-1, -3.5) {};
\node[square_node] (q2) at (0, -3.5) {};
\node[square_node] (q3) at (1, -3.5) {};
\node[square_node] (q4) at (2, -3.5) {};
\node[square_node] (q5) at (3, -3.5) {};
\node[square_node] (q6) at (4, -3.5) {};
\node[square_node] (q7) at (5, -3.5) {};
\node[square_node] (q8) at (6, -3.5) {};
\node[square_node] (q9) at (7, -3.5) {};
\node[square_node] (q10) at (8, -3.5) {};

\node[perm_node] (p1) at (0, -5.5) {$\pi_1$};
\node[perm_node] (p2) at (1.75, -5.5) {$\pi_2$};
\node[perm_node] (p3) at (3.5, -5.5) {$\pi_3$};
\node[perm_node] (p4) at (5.25, -5.5) {$\pi_4$};
\node[perm_node] (p5) at (7, -5.5) {$\pi_5$};

\node[circle_alt_node] (e1) at (0, -6.5) {SISO};

\node[circle_node2] (e11) at (-0.3, -7.5) {};
\node[circle_node2] (e12) at (-0.1167, -7.5) {};
\node[circle_node2] (e13) at (0.0667, -7.5) {};
\node[circle_node2] (e14) at (0.25, -7.5) {};

\node[circle_alt_node] (e2) at (1.75, -6.5) {SISO};

\node[circle_node2] (e21) at (1.45, -7.5) {};
\node[circle_node2] (e22) at (1.6333, -7.5) {};
\node[circle_node2] (e23) at (1.8167, -7.5) {};
\node[circle_node2] (e24) at (2, -7.5) {};

\node[circle_alt_node] (e3) at (3.5, -6.5) {SISO};

\node[circle_node2] (e31) at (3.2000, -7.5) {};
\node[circle_node2] (e32) at (3.3833 , -7.5) {};
\node[circle_node2] (e33) at (3.5667, -7.5) {};
\node[circle_node2] (e34) at (3.7500, -7.5) {};

\node[circle_alt_node] (e4) at (5.25, -6.5) {SISO};

\node[circle_node2] (e41) at (4.9500, -7.5) {};
\node[circle_node2] (e42) at (5.1333 , -7.5) {};
\node[circle_node2] (e43) at (5.3167, -7.5) {};
\node[circle_node2] (e44) at ( 5.5000, -7.5) {};

\node[circle_alt_node] (e5) at (7.0, -6.5) {SISO};

\node[circle_node2] (e51) at (6.7000, -7.5) {};
\node[circle_node2] (e52) at (6.8833 , -7.5) {};
\node[circle_node2] (e53) at (7.0667, -7.5) {};
\node[circle_node2] (e54) at (7.2500, -7.5) {};

\draw(p1.-45)--(e1.55);
\draw(p1.-74)--(e1.82);
\draw(p1.-110)--(e1.102);
\draw(p1.223)--(e1.128);

\draw(e1.223)--(e11);
\draw(e1.-105)--(e12);
\draw(e1.-79)--(e13);
\draw(e1.-50)--(e14);

\draw(p2.-45)--(e2.55);
\draw(p2.-74)--(e2.82);
\draw(p2.-110)--(e2.102);
\draw(p2.223)--(e2.128);

\draw(e2.223)--(e21);
\draw(e2.-105)--(e22);
\draw(e2.-79)--(e23);
\draw(e2.-50)--(e24);

\draw(p3.-45)--(e3.55);
\draw(p3.-74)--(e3.82);
\draw(p3.-110)--(e3.102);
\draw(p3.223)--(e3.128);

\draw(e3.223)--(e31);
\draw(e3.-105)--(e32);
\draw(e3.-79)--(e33);
\draw(e3.-50)--(e34);

\draw(p4.-45)--(e4.55);
\draw(p4.-74)--(e4.82);
\draw(p4.-110)--(e4.102);
\draw(p4.223)--(e4.128);

\draw(e4.223)--(e41);
\draw(e4.-105)--(e42);
\draw(e4.-79)--(e43);
\draw(e4.-50)--(e44);

\draw(p5.-45)--(e5.55);
\draw(p5.-74)--(e5.82);
\draw(p5.-110)--(e5.102);
\draw(p5.223)--(e5.128);

\draw(e5.223)--(e51);
\draw(e5.-105)--(e52);
\draw(e5.-79)--(e53);
\draw(e5.-50)--(e54);

\draw(b1)--(q1);
\draw(b2)--(q2);
\draw(b3)--(q3);
\draw(b4)--(q4);
\draw(b5)--(q5);
\draw(b6)--(q6);
\draw(b7)--(q7);
\draw(b8)--(q8);
\draw(b9)--(q9);
\draw(b10)--(q10);

\draw(b1)--(c1);
\draw(b1)--(c3);
\draw(b2)--(c2);
\draw(b2)--(c4);
\draw(b3)--(c3);
\draw(b3)--(c5);
\draw(b4)--(c4);
\draw(b4)--(c1);
\draw(b5)--(c5);
\draw(b5)--(c2);
\draw(b6)--(c3);
\draw(b6)--(c4);
\draw(b7)--(c4);
\draw(b7)--(c5);
\draw(b8)--(c5);
\draw(b8)--(c1);
\draw(b9)--(c1);
\draw(b9)--(c2);
\draw(b10)--(c2);
\draw(b10)--(c3);

\draw(p1.45)--(q1);
\draw(p1.110)--(q4);
\draw(p3.110)--(q1);
\draw(p2.110)--(q2);
\draw(p4.74)--(q2);
\draw(p3.74)--(q3);
\draw(p5.110)--(q8);
\draw(p4.45)--(q6);
\draw(p5.45)--(q7);
\draw(p2.74)--(q5);

\draw(p5.45)--(q5);
\draw(p2.-223)--(q9);
\draw(p3.45)--(q6);
\draw(p4.-223)--(q7);
\draw(p4.45)--(q4);
\draw(p5.-223)--(q3);
\draw(p1.-223)--(q8);
\draw(p1.74)--(q9);
\draw(p2.45)--(q10);
\draw(p3.-223)--(q10);

\end{tikzpicture}

%% file: main.bbl
\begin{thebibliography}{10}

\bibitem{higgott_improved_2023}
O.~Higgott, T.~C. Bohdanowicz, A.~Kubica, S.~T. Flammia, and E.~T. Campbell, ``Improved {Decoding} of {Circuit} {Noise} and {Fragile} {Boundaries} of {Tailored} {Surface} {Codes},'' {\em Physical Review X}, vol.~13, p.~031007, July 2023.

\bibitem{gong2024lowlatencyiterativedecodingqldpc}
A.~Gong, S.~Cammerer, and J.~M. Renes, ``{Toward Low-latency Iterative Decoding of QLDPC Codes Under Circuit-Level Noise},'' {\em arXiv preprint arXiv:2403.18901}, 2024.

\bibitem{kuo2024faulttolerantbeliefpropagationpractical}
K.-Y. Kuo and C.-Y. Lai, ``{Fault-Tolerant Belief Propagation for Practical Quantum Memory},'' {\em arXiv preprint arXiv:2409.18689}, 2024.

\bibitem{wolanski2025ambiguityclusteringaccurateefficient}
S.~Wolanski and B.~Barber, ``{Ambiguity Clustering: an Accurate and Efficient Decoder for qLDPC codes},'' {\em arXiv preprint arXiv:2406.14527}.

\bibitem{turbo_eq}
M.~Tüchler and A.~C. Singer, ``{Turbo Equalization: An Overview},'' {\em IEEE Transactions on Information Theory}, vol.~57, no.~2, pp.~920--952, 2011.

\bibitem{viterbi}
A.~Viterbi, ``{Error Bounds for Convolutional Codes and an Asymptotically Optimum Decoding Algorithm},'' {\em IEEE Transactions on Information Theory}, vol.~13, no.~2, pp.~260--269, 1967.

\bibitem{bcjr}
L.~Bahl, J.~Cocke, F.~Jelinek, and J.~Raviv, ``{Optimal Decoding of Linear Codes for Minimizing Symbol Error Rate (Corresp.)},'' {\em IEEE Transactions on Information Theory}, vol.~20, no.~2, pp.~284--287, 1974.

\bibitem{siegel_isi}
M.~H. Taghavi and P.~H. Siegel, ``{Graph-Based Decoding in the Presence of ISI},'' {\em IEEE Transactions on Information Theory}, vol.~57, no.~4, pp.~2188--2202, 2011.

\bibitem{siegel_isi2}
B.~Kurkoski, P.~Siegel, and J.~Wolf, ``{Joint Message-Passing Decoding of LDPC Codes and Partial-Response Channels},'' {\em IEEE Transactions on Information Theory}, vol.~48, no.~6, pp.~1410--1422, 2002.

\bibitem{calderbank_good_1996}
A.~R. Calderbank and P.~W. Shor, ``{Good Quantum Error-Correcting Codes Exist},'' {\em Physical Review A}, vol.~54, pp.~1098--1105, Aug. 1996.

\bibitem{bravyi2024high}
S.~Bravyi, A.~W. Cross, J.~M. Gambetta, D.~Maslov, P.~Rall, and T.~J. Yoder, ``High-{T}hreshold and {L}ow-{O}verhead {F}ault-{T}olerant {Q}uantum {M}emory,'' {\em Nature}, vol.~627, no.~8005, pp.~778--782, 2024.

\bibitem{kay2023tutorialquantikzpackage}
A.~Kay, ``Tutorial on the quantikz package,'' {\em arXiv preprint arXiv:1809.03842}, 2023.

\bibitem{dgldpc}
Y.~Wang and M.~Fossorier, ``{Doubly Generalized LDPC Codes},'' in {\em 2006 IEEE International Symposium on Information Theory}, pp.~669--673, 2006.

\bibitem{nms_decoder}
J.~Chen, A.~Dholakia, E.~Eleftheriou, M.~Fossorier, and X.-Y. Hu, ``{Reduced-Complexity Decoding of LDPC Codes},'' {\em IEEE Transactions on Communications}, vol.~53, no.~8, pp.~1288--1299, 2005.

\bibitem{chytas2025enhancedminsumdecodingquantum}
D.~Chytas, N.~Raveendran, and B.~Vasi\'c, ``{Enhanced Min-Sum Decoding of Quantum Codes Using Previous Iteration Dynamics},'' {\em arXiv preprint arXiv:2501.05021}, 2025.

\bibitem{gidney2021stim}
C.~Gidney, ``{Stim: a Fast Stabilizer Circuit Simulator},'' {\em {Quantum}}, vol.~5, p.~497, July 2021.

\end{thebibliography}
